\documentclass[prl,floatfix,10pt,amsmath,twocolumn,superscriptaddress,amsfonts,amssymb,aps]{revtex4-2}
\pdfoutput=1
\usepackage[T1]{fontenc}\usepackage[latin1]{inputenc}
\usepackage{dcolumn,graphicx,color} 
\graphicspath{{Figures/}}
\usepackage{sidecap}

\usepackage[colorlinks,plainpages=false,linkcolor=blue,urlcolor=blue,citecolor=blue,pdfpagemode=UseNone,pdfstartview=FitBH]{hyperref}
\usepackage{orcidlink} 

\makeatletter
\newcommand{\bibstyle@supplement}{\bibpunct[, ]{[S}{]}{;}{n}{,}{, S\hspace{-2.1pt}}%
\gdef\bibnumfmt##1{[S##1]}}
\makeatother
\newcommand{\dcu}{\ensuremath{d_{\text{Cu--Cu}}}}

\begin{document}

\makeatletter\renewcommand{\ps@plain}{%
\def\@evenhead{\hfill\itshape\rightmark}%
\def\@oddhead{\itshape\leftmark\hfill}%
\renewcommand{\@evenfoot}{\hfill\small{--~\thepage~--}\hfill}%
\renewcommand{\@oddfoot}{\hfill\small{--~\thepage~--}\hfill}%
}\makeatother\pagestyle{plain}

\title{Vestiges of magnon fractionalization in the quasi-two-dimensional quantum magnet\\ rouaite Cu$_\text{2}$(NO$_\text{3}$)(OH)$_\text{3}$ with alternating ferro- and antiferromagnetic {\itshape S}=$\frac{\text{1}}{\text{2}}$ chains}

\author{Darren~C.~Peets\,\orcidlink{0000-0002-5456-574X}}
\email[E-mail: ]{darren.peets@tu-dresden.de}
\affiliation{Institut f\"ur Festk\"orper- und Materialphysik, Technische Universit\"at Dresden, 01069 Dresden, Germany}

\author{Satoshi Nishimoto\,\orcidlink{0000-0001-9768-7192}}
\affiliation{Institut f\"ur Theoretische Physik, Technische Universit\"at Dresden, 01069 Dresden, Germany}
\affiliation{Institute for Theoretical Solid State Physics, IFW Dresden, Helmholtzstra{\ss}e 20, 01069 Dresden, Germany}

\author{Denys Y.\ Kononenko\,\orcidlink{0000-0002-4467-730X}}
\author{Oleg Janson\,\orcidlink{0000-0001-7328-5690}}
\email[E-mail: ]{o.janson@ifw-dresden.de}
\affiliation{Institute for Theoretical Solid State Physics, IFW Dresden, Helmholtzstra{\ss}e 20, 01069 Dresden, Germany}

\author{Aswathi Mannathanath Chakkingal}
\affiliation{Institut f\"ur Festk\"orper- und Materialphysik, Technische Universit\"at Dresden, 01069 Dresden, Germany}

\author{Anton~A.~Kulbakov\,\orcidlink{0000-0001-9676-8310}}
\affiliation{Institut f\"ur Festk\"orper- und Materialphysik, Technische Universit\"at Dresden, 01069 Dresden, Germany}

\author{Kaushick~K.~Parui\,\orcidlink{0000-0002-6584-7830}}
\affiliation{Institut f\"ur Festk\"orper- und Materialphysik, Technische Universit\"at Dresden, 01069 Dresden, Germany}

\author{Harald O.\ Jeschke\,\orcidlink{0000-0002-8091-7024}}
\affiliation{Research Institute for Interdisciplinary Science, Okayama University, Okayama 700-8530, Japan}

\author{Ellen H{\"a}u{\ss}ler\,\orcidlink{0000-0001-5620-836X}}
\affiliation{Anorganische Chemie II, Technische Universit{\"a}t Dresden, 01062 Dresden, Germany}

\author{James P.\ Tellam\,\orcidlink{0000-0001-9243-1323}}
\author{J.~Ross Stewart\,\orcidlink{0000-0003-0053-0178}}
\affiliation{ISIS Neutron and Muon Source, Rutherford Appleton Laboratory, Didcot OX11 0QX, UK}

\author{Dmytro S. Inosov\,\orcidlink{0000-0003-0639-2503}}\email[E-mail: \smallskip]{dmytro.inosov@tu-dresden.de}
\affiliation{Institut f\"ur Festk\"orper- und Materialphysik, Technische Universit\"at Dresden, 01069 Dresden, Germany}
\affiliation{W\"urzburg-Dresden Cluster of Excellence on Complexity and Topology in Quantum Matter\,---\,\textit{ct.qmat}, TU Dresden, 01069 Dresden, Germany}

\begin{abstract}
Ferro- and antiferromagnetic quantum spin chains host qualitatively distinct elementary excitations\,---\,magnons ($S=1$, bosonic quasiparticles) and spinons ($S=\frac{1}{2}$, fractionalized quasiparticles), respectively. Weakly coupled networks of both chain types therefore provide an ideal platform for exploring spinon-magnon coupling and the hybridization of sharp magnon modes with spinon continua. Here we investigate the spin-excitation spectrum of the frustrated quantum magnet rouaite, Cu$_2$(NO$_3$)(OH)$_3$\,---\,a model realization of weakly coupled ferro- and antiferromagnetic Cu$^{2+}$ spin-$\frac{1}{2}$ chains. The frustrated interchain interactions, while weaker than the dominant intrachain exchange, are sufficient to stabilize a long-range-ordered state that is not reproduced by classical calculations. Neutron spectroscopy reveals a spectrum of well-defined spin waves associated with both types of chains, with clear signatures of hybridization but no observable spinon continuum. Certain magnon branches resulting from this hybridization exhibit substantial broadening near the zone center. We attribute this to vestiges of spinon fractionalization persisting partway through the dimensional crossover from isolated quantum spin chains to a 2D network of frustrated spins.
\end{abstract}

\maketitle

\paragraph{Introduction.} Ferromagnetic (FM) and antiferromagnetic (AFM) quantum spin chains support fundamentally different elementary magnetic excitations: magnons and spinons, respectively. Magnons are spin-1 quasiparticles that form sharp dispersive spin-wave branches, whereas spinons are fractionalized \mbox{spin-$\frac{1}{2}$} excitations that appear as broad continua in spectroscopic probes, reflecting the fact that most such probes couple to multi-spinon excitations rather than to individual spinons. Systems containing both types of chains therefore provide a natural setting to investigate the interplay between these distinct quasiparticles.

Alternating FM and AFM spin chains are a motif common to a class of layered copper hydroxysalt minerals exemplified by $\alpha$-Cu$_2$(OH)$_3$Cl (botallackite)~\cite{ZhengMori05, Inosov18}, its synthetic analogues Cu$_2$(OH)$_3$Br~\mbox{\cite{Oswald1961,ZhengYamashita09, ZhangZhao20}} and Cu$_2$(OH)$_3$I~\cite{Oswald1961,ZhengYamashita09}, Cu$_2$(OH)$_3$NO$_3$ (rouaite)~\cite{ChakkingalKulbakov24, WulferdingIslam25, NiwataKumar26}, and possibly in the structurally related but less-studied 
layered copper-hydroxide frameworks Cu$_2$(OH)$_3$(\text{alkanecarboxylate}) and related systems~\cite{FujitaAwaga96,RabuRouba96,FujitaAwaga99,Rogez2011}. In particular, the existence of a spinon continuum hybridizing with the magnon modes was recently proposed in Br-botallackite based on inelastic neutron scattering (INS) measurements~\cite{ZhangZhao20} and later in rouaite from Raman spectra~\cite{WulferdingIslam25}.

\begin{figure}[b!]
\includegraphics[width=\linewidth]{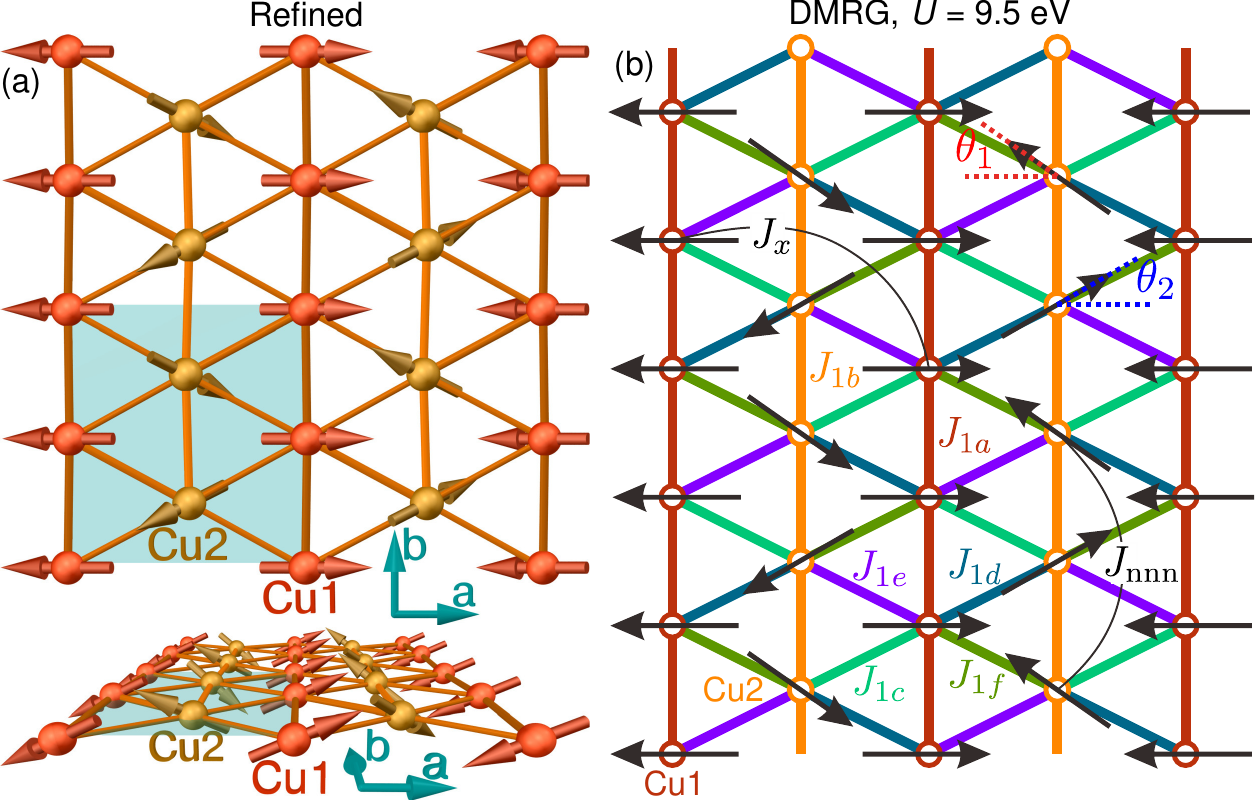}\vspace{-1em}
\caption{Magnetic ground state of rouaite from (a)~low-temperature neutron diffraction confirmed by NMR measurements and (b)~symmetry-breaking DMRG for $U=9.5$~eV couplings, with exchange paths and canting angles indicated.
}\label{Fig:Magnetic_Structures}
\end{figure}

Rouaite crystallizes in the monoclinic $P2_1$ space group and orders magnetically below the N\'eel temperature $T_\text{N}=7.24$~K~\cite{Linder1995,Kikuchi2018,ChakkingalKulbakov24}. The FM and AFM spin chains comprise the crystallographically inequivalent Cu1 and Cu2 sites, respectively, and run along $\mathbf{b}$. They alternate along $\mathbf{a}$ to form weakly coupled 2D planes. According to both neutron diffraction~\cite{ChakkingalKulbakov24} and nuclear magnetic resonance (NMR)~\cite{NiwataKumar26}, the magnetic ground state of rouaite adopts a commensurate non-coplanar long-range magnetic order with propagation vector $\mathbf{q}_\text{m}=(\frac{1}{2}\,0\,0)$ [Fig.~\ref{Fig:Magnetic_Structures}(a)]. The Cu1 spins point approximately along the $\mathbf{a}$ direction with a non-negligible out-of-plane canting. Consecutive FM chains are antialigned, resulting in a compensated AFM structure without net magnetization. In contrast to botallackite, the Cu2 spins possess an additional canting along the $\mathbf{b}$ direction, resulting in a net magnetization that alternates from one AFM chain to the next.

According to our earlier study~\cite{ChakkingalKulbakov24}, this state is rather fragile and undergoes a first-order transition to an incommensurate phase with a similar in-plane structure plus an additional $c$-axis modulation either upon warming above $T_1=4.7$~K or in a weak magnetic field orthogonal to the chains. At even higher fields perpendicular to the chains (on the order of 2~T), the ferromagnetic chains undergo another discontinuous spin-flop transition into an uncompensated state. Further increasing the magnetic field continuously changes the canting angle until the FM chains become fully polarized around 20~T, resulting in an effective decoupling of the AFM chains and the formation of a putative field-induced Tomonaga-Luttinger liquid, as also now proposed in the Br analogue~\cite{Khatua2608.04096}.

\begin{figure*}[t!]
\begin{center}
\includegraphics[width=\linewidth]{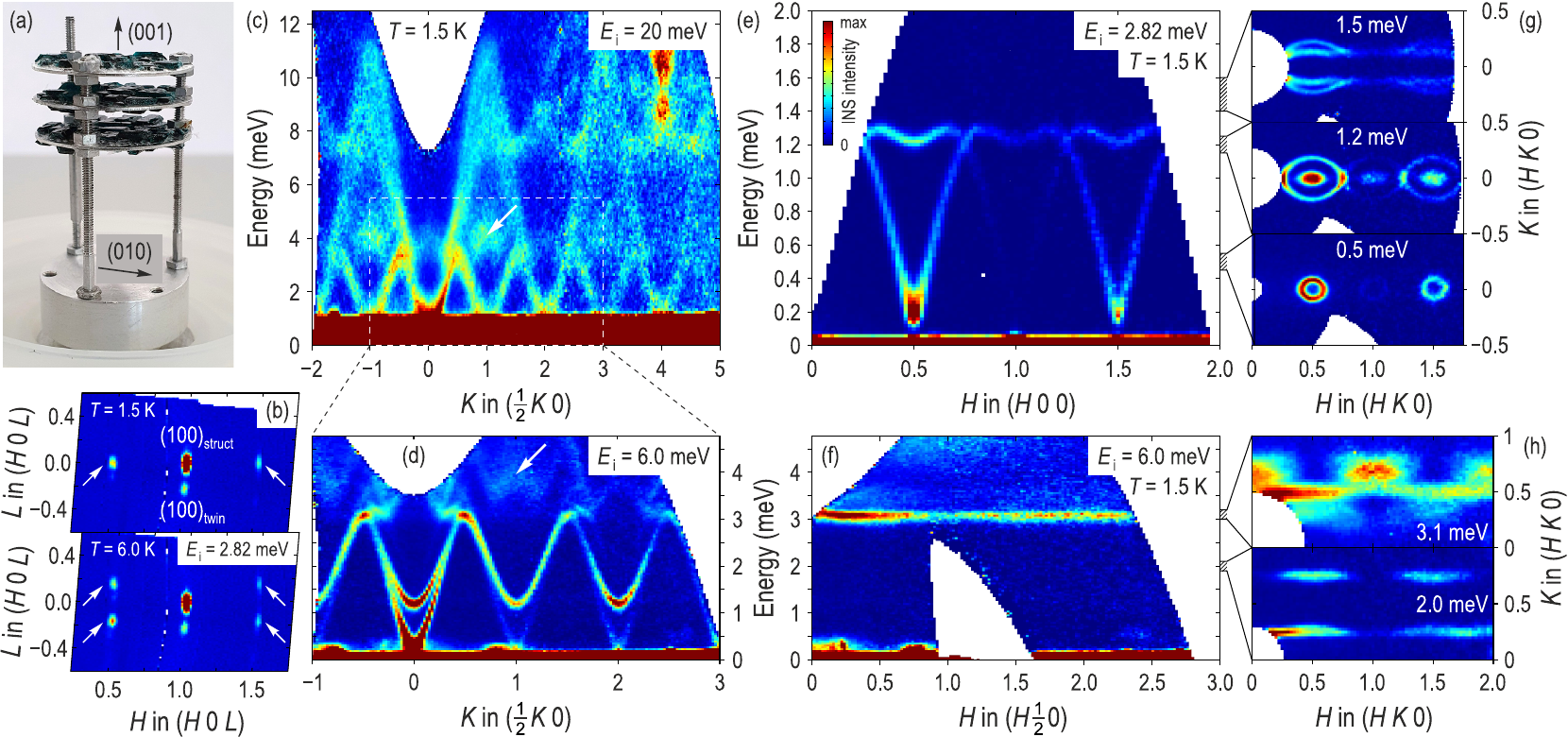}\vspace{-1em}
\end{center}
\caption{Summary of single-crystal INS data on rouaite. (a)~Coaligned single-crystal mosaic sample of deuterated Cu$_2$(NO$_3$)(OD)$_3$ used in the INS experiment on \textsc{Let}, ISIS. (b)~Elastic ($E=0$) cuts through the TOF dataset, showing the magnetic Bragg peaks in the commensurate ($T=1.5$~K) and incommensurate ($T=6.0$~K) phases. (c,\,d)~Momentum-energy cuts along $(\frac12K0)$ measured with $E_\text{i}=20$ and 6.0 meV, respectively. (e,\,f)~Momentum-energy cuts along the $(H00)$ and $(H\frac{1}{2}0)$ directions, respectively. (g,\,h)~Constant-energy cuts in the $(HK0)$ plane at several representative energies as indicated in each panel. The data are integrated within the hatched energy ranges shown on the left.}\label{Fig:INS}
\end{figure*}

\begin{table}[b]\caption{\label{tab:js}Magnetic exchange interactions (in K) determined using total-energy GGA+$U$ calculations with $U$ = 9.5~eV. Interatomic distances and Cu(1)/Cu(2) notation refer to the experimental crystal structure described in Tables I and IV of Ref.~\cite{ChakkingalKulbakov24}. Uncertainties determined from bootstrapping least-squares fitting pertain to the last digit.
}
\begin{ruledtabular}
\begin{tabular}{ldcd}
Exchange & \multicolumn{1}{r}{\dcu} & Pathway & \multicolumn{1}{c}{$J$ (K)} \\ \hline
$J_{1a}$         & 3.041 & Cu(1)--Cu(1) & -26.7(1) \\
$J_{1b}$         & 3.042 & Cu(2)--Cu(2) &  23.1(1) \\
$J_{1d}$         & 3.136 & Cu(1)--Cu(2) &   0.9(1) \\
$J_{1c}$         & 3.142 & Cu(1)--Cu(2) &  -1.8(1) \\
$J_{1f}$         & 3.222 & Cu(1)--Cu(2) &  12.6(1) \\
$J_{1e}$         & 3.246 & Cu(1)--Cu(2) &  21.5(1) \\
$J_{\text{nnn}}$ & 6.080 & Cu(2)--Cu(2) &  20.0(4) \\
$J_{x}$          & 8.266 & Cu(1)--Cu(1) &  -0.8(1) \\
\end{tabular}
\end{ruledtabular}
\end{table}

\paragraph{Density functional theory (DFT) calculations.} In our DFT analysis, details of which are in the Supplemental Material\,\cite{SupplementalMaterial}, we consider eight exchange interactions: nearest-neighbor exchanges in both chains ($J_{1a}$ and $J_{1b}$); four inequivalent short-range (\dcu~$\leq$~3.5\,\AA) interactions between the corner- and edge-sharing chains\,---\,$J_{1c}$ to $J_{1f}$; the second-neighbor interaction in the edge-sharing chain $J_{\text{nnn}}$; as well as the long-range interchain exchange $J_x$ connecting two nearly coplanar CuO$_4$ units. Numerical estimates for these eight interactions are provided in Table~\ref{tab:js}.  The nearest-neighbor exchanges $J_{1a}$ and $J_{1b}$ are both sizable and have opposite sign: antiferromagnetic in the edge-sharing chain and ferromagnetic in the corner-sharing chain. For the latter, the sizable noncoplanarity of the neighboring CuO$_4$ units drastically suppresses the antiferromagnetic superexchange contribution. The four interchain interactions split into two groups: while $J_{1f}$ and especially $J_{1e}$ are relevant and comparable to $|J_{1a}|$, the short-range $J_{1c}$ and $J_{1d}$ are significantly smaller and can be neglected in the minimal model. The frustrated second-neighbor interaction within the edge-sharing chain, $J_{\text{nnn}}$, is again sizable and
comparable to $J_{1e}$. Finally, the long-range interchain exchange $J_x$ is very
weak. We note that the large number of computed configurations (48) and small
uncertainties (see also the last column of Table~\ref{tab:configs}) obtained in
our bootstrapping procedure practically guarantee the completeness of the spin
model\,---\,all further exchanges beyond the eight terms listed in
Table~\ref{tab:js} do not exceed a few percent of the leading term ($J_{1a}$).
Although our computational method is restricted to collinear magnetism, the
lowest-energy configuration (top row in Table~\ref{tab:configs}) is very
similar to the experimental ground state: it features ferromagnetic
corner-sharing chains, with an alternation of ``up'' and ``down'' chains if we
move along $a$, and antiferromagnetic edge-sharing chains that show a similar
alternation (see Fig.~\ref{fig:collinear_gs}).

\paragraph{Inelastic neutron scattering.} We performed INS measurements of the spin-excitation spectrum of rouaite on the cold-neutron time-of-flight (TOF) spectrometer \textsc{Let} at the ISIS neutron source, UK. To avoid the high incoherent-scattering background from hydrogen, we used deuterated single crystals of rouaite, Cu$_2$(NO$_3$)(OD)$_3$, which were grown hydrothermally as described elsewhere~\cite{ChakkingalKulbakov24}. Multiple crystals were coaligned using x-ray Laue into a mosaic sample with a total mass of 1.7~g, shown in Fig.~\ref{Fig:INS}\,(a). This sample was loaded with the $\textbf{c}^\ast$ axis vertical in a standard ``orange'' cryostat with a base temperature of 1.5~K and measured in the repetition-rate multiplication (``multirep'') mode with three incident neutron energies $E_\text{i}=20$, 6.0, and 2.82~meV. Representative cuts through the TOF dataset are shown in Fig.~\ref{Fig:INS}. Since the out-of-plane magnon bandwidth does not exceed 0.02~meV (see Fig.~\ref{Fig:LET_Gap}~\cite{SupplementalMaterial}), we treat the data as two-dimensional and integrate the INS intensity over the full accessible out-of-plane range to maximize the signal-to-noise ratio.

Figure~\ref{Fig:INS}\,(b) shows part of the elastic ($E=0$) cut through the TOF datasets measured at 1.5 and 6.0~K that covers the $(100)$ structural reflection and two magnetic Bragg peaks at $(\frac{1}{2}00)$ and $(\frac{3}{2}00)$ (emphasized with white arrows). A much weaker $(100)_\text{twin}$ reflection from the monoclinic twin domain is visible, indicating a twin-domain fraction of about 9\% in the sample mosaic whose contribution to the magnetic scattering is negligible. Above $T_1$, these peaks are split along $\textbf{c}$ in the incommensurate phase~\cite{ChakkingalKulbakov24}.

A bird's-eye view of the spin-wave dispersions along the chain direction, $(\frac{1}{2}K0)$, is presented in Fig.~\ref{Fig:INS}\,(c). One can see two qualitatively distinct sets of spin-wave modes with bandwidths of approximately 3 and 11~meV and different periodicity. The band extending to 11~meV has minima at even $K$ and is therefore naturally ascribed to the FM chains. This magnon mode shows a marginally resolved splitting near the top of the band. The second band reaching 3~meV has minima at every integer $K$ and must therefore stem from the AFM chains whose periodicity along the chains is doubled in direct space. In addition, we see a diffuse ``cloud'' of intensity centered around \mbox{$(\frac{1}{2}\pm\kern-1.5pt1\,0)$} at 4~meV (white arrow), seemingly disconnected from both bands. Its intensity decreases with $|\mathbf{Q}|$ in accordance with the magnetic form factor, indicating its magnetic origin. By contrast, additional features that appear above 7~meV must originate from optical phonons, as their intensity increases toward larger~$|\mathbf{Q}|$.

A zoomed-in view of the same cut, shown in Fig.~\ref{Fig:INS}\,(d), further reveals that the low-energy magnon band is split into two modes. These modes remain sharp (resolution limited) throughout reciprocal space, in sharp contrast to the 4-meV feature (white arrow) that retains its substantial broadening despite the improved resolution at this $E_i$. We also observe a complex pattern emerging near the top of the band above 3~meV, possibly arising from the hybridization of the AFM and FM magnon bands. To understand the origin of the spin-wave splitting, we examine the cut passing through the ordering vector in the orthogonal in-plane direction, $(H00)$, taken from the $E_\text{i}=2.82$~meV dataset in Fig.~\ref{Fig:INS}\,(e). We see that both bands have the same dispersion but are shifted by half a reciprocal lattice vector along the $H$ direction. This reflects the doubling of the magnetic unit cell along the $\mathbf{a}$ axis that results from the alternating arrangement of the spin chains. With the higher energy resolution, we can now also resolve a 0.2-meV anisotropy gap in the low-energy magnon mode.

\begin{figure}[b!]
\begin{center}
\includegraphics[width=\linewidth]{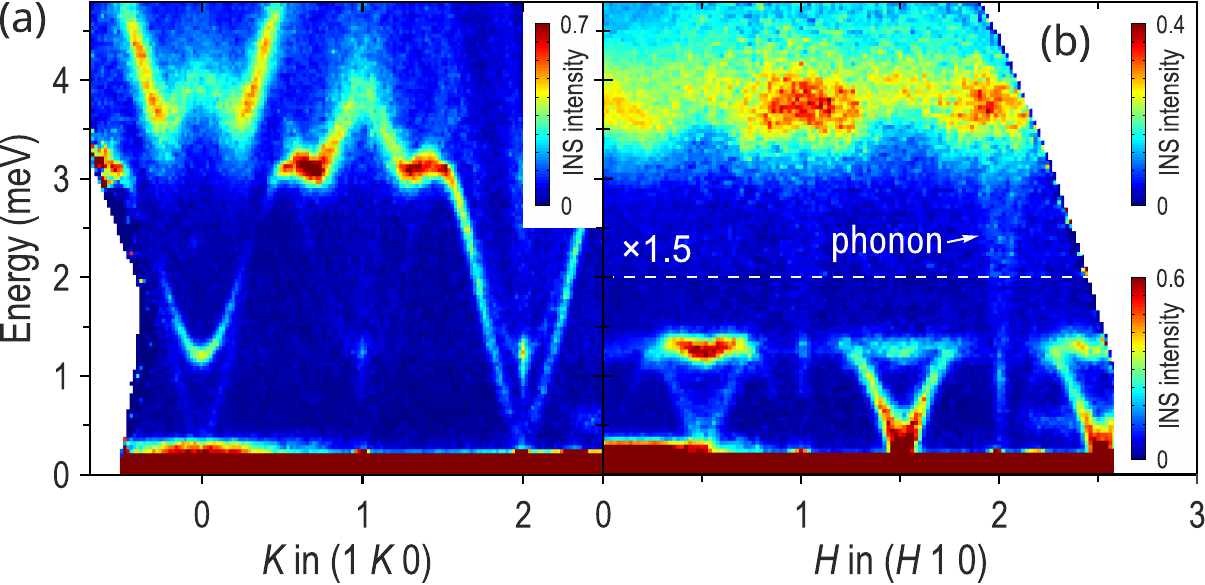}\vspace{-1em}
\end{center}
\caption{Orthogonal cuts along (a)~$(1K0)$ and (b)~$(H10)$ that highlight the dispersion of the broad 4-meV feature at the zone center $(110)$ and its connection to the spin-wave modes.\vspace{-4pt}}\label{Fig:1K0+H10}
\end{figure}

The magnon bandwidth along the $\mathbf{a}$ direction amounts to 1.3~meV, an order of magnitude smaller than the bandwidth along $\mathbf{b}$. This suggests a strong hierarchy of energy scales, with interchain interactions an order of magnitude weaker than the dominant intrachain exchanges. Nevertheless, the low-energy spin waves clearly cannot be described as one-dimensional (1D). This is especially clear from Fig.~\ref{Fig:INS}\,(g), which shows constant-energy cuts through the $E_\text{i}=2.82$~meV dataset, revealing clear dispersion cones with a notable dispersion along both directions up to at least 1.5~meV. Above this energy, the $H$-dependence remains visible as strong intensity modulations in the $E_\text{i}=6.0$~meV dataset, see Fig.~\ref{Fig:INS}\,(h). The only cut through reciprocal space where the dispersion along $H$ vanishes is the $(H\frac{1}{2}0)$ cut shown in Fig.~\ref{Fig:INS}\,(f), which passes through the top of the low-energy magnon band. It is worth noting that exactly this cut was used by the authors of Ref.~\cite{ZhangZhao20} to argue that Br-botallackite has 1D spin waves which lack dispersion along both the $H$ and $L$ directions, thereby justifying the integration of their INS dataset along $H$. Based on our present results and rouaite's structural similarity to botallackite, we consider that such integration can introduce additional broadening of the spin-wave modes which may be mistaken for a spinon continuum. It therefore remains unclear to what extent the reported spinon continuum in Br-botallackite may have been influenced by this integration procedure. For a result of such integration in rouaite, see Fig.~\ref{Fig:LET_H-integrated} in the Supplemental Material~\cite{SupplementalMaterial}.

Two orthogonal momentum-energy cuts along $(1K0)$ and $(H10)$ that cross the broad 4-meV feature are presented in Fig.~\ref{Fig:1K0+H10} (see also Fig.~\ref{Fig:LET_Spectra_H00} in the Supplemental Material~\cite{SupplementalMaterial}). Figure~\ref{Fig:1K0+H10}\,(a) shows that at zone centers with even $K$ this feature disperses upward in energy and connects to the FM spin-wave branch, whereas at odd $K$ it disperses downward and connects to the AFM mode. It must therefore result from hybridization of spin waves associated with both chains. As noted earlier, odd and even $K$ are equivalent for the AFM chain but correspond to the top and bottom of the FM magnon band, respectively. Near 3~meV, where the two bands approach one another, a complex hybridization pattern emerges, producing multiple weak features visible in Fig.~\ref{Fig:INS}\,(d). Among them is a third downward-dispersing magnon branch that reaches the zone center at 1.8~meV with vanishing intensity. This mode could account for the magnon feature observed in Raman spectroscopy~\cite{WulferdingIslam25}, as also discussed in Sec.~S3 of the Supplemental Material~\cite{SupplementalMaterial}. As we show in Fig.~\ref{Fig:1K0+H10}\,(b), the broad 4-meV feature also exhibits a weak dispersion and an intensity modulation orthogonal to the chains. Here the substantial broadness of this feature relative to the sharp, resolution-limited low-energy magnon modes becomes most apparent.

\paragraph{Spin-dynamical calculations.}

\begin{figure*}[t!]
\includegraphics[width=0.9\linewidth]{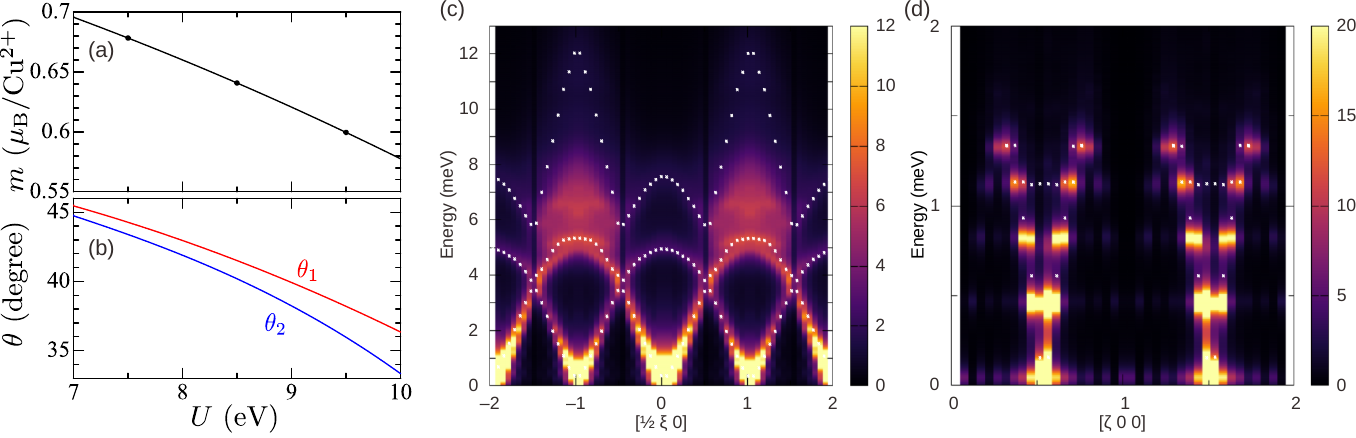}\vspace{-1em}
\caption{
(a)~Ordered moment per Cu$^{2+}$ ($g=2$) as a function of the
DFT+$U$ parameter $U$, obtained by symmetry-breaking DMRG.
(b)~Canting angles $\theta_1$ and $\theta_2$ of the Cu2 spins
[defined in Fig.~1(b)], obtained by ED with a pinning field.
(c),(d)~$S(\mathbf{Q},\omega)$ calculated by DDMRG for $U=9.5$~eV along (c)~$(\frac{1}{2}\,\xi\,0)$ and (d)~$(\zeta\,0\,0)$, with Lorentzian broadening $\eta=0.05J_{1b}\simeq0.22$~meV and $0.0125J_{1b}\simeq0.054$~meV, respectively; cf.\ the INS data in Fig.~\ref{Fig:INS}. Crosses mark the extracted excitation dispersions (see~Supplemental Material\,\cite{SupplementalMaterial}). Color scales are saturated to emphasize weak features.
}\label{Fig:DMRG}
\end{figure*}

To verify that the DFT-derived exchange couplings reproduce the observed magnetic order, we studied the corresponding spin-$\frac{1}{2}$ Heisenberg model using the density-matrix renormalization group (DMRG) method~\cite{White92}. The ground state was obtained on a $17\times8$ cluster with open boundary conditions, retaining up to 2400 density-matrix eigenstates to keep the discarded weight below $10^{-6}$. Allowing for spontaneous symmetry breaking, the calculation converges to the commensurate noncoplanar order described above, in which the Cu2 spins cant alternately by two inequivalent angles $\theta_1$ and $\theta_2$ [Fig.~\ref{Fig:DMRG}(b)]. Remarkably, this magnetic order\,---\,too fragile to be captured by classical energy minimization\,---\,is robustly reproduced over the entire realistic range of the Coulomb parameter $U$ used in DFT+$U$ calculations ($U=7$--$10$~eV), once quantum fluctuations are fully taken into account, demonstrating that the DFT+$U$-derived couplings contain the essential ingredients of the magnetism in rouaite. The $U$ dependence enters only quantitatively: the ordered moment per Cu$^{2+}$ decreases monotonically from $0.70$ to $0.58\,\mu_\mathrm{B}$ (assuming $g=2$) [Fig.~\ref{Fig:DMRG}(a)], since the intrachain coupling $J_{1b}$ weakens much faster than the frustrated interchain couplings $J_{1c}$--$J_{1f}$, enhancing the relative frustration and hence quantum fluctuations. The canting angles, determined independently by exact diagonalization of a $6\times4$ cluster with periodic boundary conditions using local pinning fields\,\cite{SupplementalMaterial}, likewise decrease with $U$ [Fig.~\ref{Fig:DMRG}(b)]. As shown below, the magnon bandwidths observed by INS pin down the overall energy scale to $U\simeq9.5$~eV; at this value, the calculated ordered moment is $m\simeq0.60\,\mu_{\rm B}$ and the canting angles are $(\theta_1,\theta_2)=(38^\circ,36^\circ)$. These values reproduce the noncollinear character of the experimental magnetic structure, although both the moment and the canting angles are larger than the neutron-diffraction estimates, $m_{\rm exp}=0.47\,\mu_{\rm B}$ and $\theta_{\rm exp}=22^\circ$\footnote{It was not possible to separately refine these two canting angles from the neutron diffraction data.}. The quantitative differences may partly reflect finite-size effects, finite measurement temperatures, and, for the canting angles, the use of local pinning fields. 

The dynamical spin structure factor $S(\mathbf{Q},\omega)$ was calculated by the dynamical DMRG (DDMRG) method~\cite{Jeckelmann02} on $17\times4$ and $4\times20$ clusters for momenta along and perpendicular to the chains, respectively [Figs.~\ref{Fig:DMRG}(c) and \ref{Fig:DMRG}(d)]. The measured bandwidths of the FM and AFM magnon branches ($\sim$11 and $\sim$3~meV) are best reproduced by the $U=9.5$~eV couplings, which fixes $U$ more directly than the static quantities: the bandwidths scale with the exchange couplings themselves, whereas the moment and canting angles depend on them only through ratios. Along $(\frac{1}{2}\,\xi\,0)$, the calculation reproduces the two qualitatively distinct sets of spin-wave modes observed by INS: the dispersive magnon band of the FM chains with minima at even $\xi$, and the low-energy band of the AFM chains with doubled periodicity, including the complex intensity pattern near the top of the latter arising from their hybridization. The calculation also yields broad continuum-like intensity at $\sim$6\,meV, somewhat higher than the experimental value of $\sim$4\,meV. Reducing the couplings between neighboring chains relative to the dominant intrachain exchange improves the agreement in its characteristic energy after matching the overall bandwidth to experiment (see Supplemental Material~\cite{SupplementalMaterial}). This suggests that these coupling ratios may be somewhat smaller than those estimated by DFT+$U$ at $U = 9.5$ eV. We note that the overall structure of this spectrum\,---\,a dome at higher energy containing diffuse intensity which is perched atop magnon-band domes of roughly half the bandwidth which contain no diffuse intensity\,---\,closely resembles that of spinon continua observed in 1D spin chains\,\cite{BeraLake17}.

Along $(\zeta\,0\,0)$, the DDMRG spectrum disperses around the ordering vector $(\frac{1}{2}\,0\,0)$, confirming the 2D character of the low-energy spin waves. Around $\zeta\simeq1$, the isotropic model yields faint but finite spectral weight following the magnon dispersion [Fig.~\ref{Fig:DMRG}(d)], considerably weaker than observed. The mechanism that transfers weight to this zone boundary is thus already contained in the isotropic model, while its enhancement in the experiment points to anisotropic interactions allowed by the $P2_1$ symmetry\,---\,consistent with the 0.2-meV anisotropy gap\,---\,which are not expected to affect the dispersions on the relevant energy scales. The overall agreement, achieved with no adjustable parameters beyond the choice of $U$, demonstrates that the isotropic Heisenberg model captures both the fragile magnetic order and the spin dynamics of rouaite.

\paragraph{Discussion and Conclusion.} We have shown that the spin-excitation spectrum of rouaite contains no spinon continuum in the conventional sense. The low-energy AFM spin waves remain sharp with no additional tails of diffuse scattering intensity. However, the spectrum also cannot be described solely by sharp magnon modes. The intrinsically broad peak centered at 4~meV, where the FM and AFM spin waves hybridize, clearly distinguishes rouaite from a classical spin system with three-dimensional long-range magnetic order. This feature likely accounts for the broad magnetic background with an onset at 3~meV that was observed earlier in the Raman spectra~\cite{WulferdingIslam25} (see section S3 in the Supplemental Material~\cite{SupplementalMaterial}).

As interchain interactions increase, spinon confinement is expected to transfer most continuum spectral weight into sharp spin-wave modes~\cite{Schulz96, EsslerTsvelik97, RaczkowskiAssaad13, BeraLake17}. At the same time, hybridization of two conventional magnon modes should produce avoided crossings between otherwise sharp branches. The substantial broadness of the 4-meV feature therefore suggests that the hybridized response retains vestiges of continuum-derived spectral weight inherited from the weakly coupled AFM chains near the avoided crossing. As a result, the intrinsic width of the hybrid modes is enhanced where the mixing is strongest. This can be understood as momentum-dependent spinon confinement: in rouaite, the weak interchain coupling appears sufficient to transfer most spinon-continuum spectral weight into sharp spin-wave modes over much of reciprocal space, while leaving the confinement incomplete near the wave vector where the AFM and FM spin waves hybridize. Equivalently, from the spin-wave side, this is analogous to a momentum-dependent magnon breakdown~\cite{OhLe13, ParkPark20}: near the avoided crossing, the hybridized magnon pole couples to continuum-derived spectral weight and loses its sharp quasiparticle character.

Rouaite and the broader botallackite family offer a unique opportunity to tune the dimensionality of a 2D plane comprising AFM and FM spin chains, to examine how spinon continua evolve into magnon bands.  The vestiges of spinon continua visible in rouaite indicate that this process is not abrupt and intermediate cases are possible, just as the materials are neither purely 1D nor 2D.

\paragraph{Data availability.} The data that support the findings of this article are openly available~\cite{LET-RB2320001}. Questions about the deposited files and requests for additional material may be directed to the corresponding authors.

\begin{acknowledgments}
This project was funded by the Deutsche Forschungsgemeinschaft (DFG, German Research Foundation) through individual grant PE~3318/2 (Project No.\ 452541981); through projects A05, B03, and C03 of the Collaborative Research Center SFB~1143 (Project No.\ 247310070); and through the W\"urzburg-Dresden Cluster of Excellence on Complexity and Topology in Quantum Materials\,---\,\textit{ct.qmat} (EXC~2147, Project No.\ 390858490). We gratefully acknowledge the UK Science and Technology Facilities Council (STFC) for access to neutron beamtime at ISIS and for the provision of sample deuteration facilities\,\cite{LET-RB2320001}. 
\end{acknowledgments}

\bibliography{rouaite}\vspace{-2pt}

\clearpage
\pagestyle{plain}

\renewcommand\thefigure{S\arabic{figure}}
\renewcommand\thetable{S\arabic{table}}
\renewcommand\theequation{S\arabic{equation}}
\renewcommand{\theHfigure}{S\arabic{figure}}
\renewcommand{\theHtable}{S\arabic{table}}
\renewcommand{\theHequation}{S\arabic{equation}}
\setcounter{page}{1}
\setcounter{figure}{0}
\setcounter{table}{0}
\setcounter{equation}{0}


\pagestyle{plain}
\makeatletter
\renewcommand{\@oddfoot}{\hfill\bf\scriptsize\textsf{S\thepage}}
\renewcommand{\@evenfoot}{\bf\scriptsize\textsf{S\thepage}\hfill}
\renewcommand{\@oddhead}{D.\hspace{0.6ex}C.\hspace{0.6ex}Peets \textit{et~al.}\hfill\Large\textsf{{Supplemental Material}}}
\renewcommand{\@evenhead}{D.\hspace{0.6ex}C.\hspace{0.6ex}Peets \textit{et~al.}\Large\textsf{{Supplemental Material}}\hfill}
\renewcommand{\fnum@figure}[1]{\figurename~\thefigure.~}

\makeatother
\setcounter{page}{1}\setcounter{figure}{0}\setcounter{table}{0}\setcounter{equation}{0}

\onecolumngrid\normalsize
\begin{center}{\vspace*{0.1pt}\large{Supplemental Material to the Letter\smallskip\\\sl\textbf{``Vestiges of magnon fractionalization in the quasi-two-dimensional quantum magnet\\ rouaite Cu$_\text{2}$(NO$_\text{3}$)(OH)$_\text{3}$ with alternating ferro- and antiferromagnetic {\itshape S}\ =\ $\frac{\text{1}}{\text{2}}$ chains''}}}\end{center}\bigskip

\twocolumngrid

\bigskip
\noindent\textbf{\it S1.~Out-of-plane dispersion and $T$-dependence of the spin gap}\vspace*{5pt}

Figure~\ref{Fig:LET_Gap} shows INS data in the out-of-plane direction $(\frac{1}{2}0L)$ that reveal a very weak dispersion with a bandwidth not exceeding 0.02\,meV. The band minimum at $(\frac{1}{2}00)$ corresponds to the anisotropy gap of 0.2~meV discussed in the main text. This gap increases to only $\sim$\,0.22~meV at the zone boundary, $(\frac12 0\frac12)$. Since this dispersion is much weaker than the gap itself, we have neglected it in the data analysis and integrated all data presented in the main paper over the entire accessible $L$ range to maximize the signal-to-noise ratio.

To resolve the spin gap clearly at elevated temperatures, we have measured additional datasets at $T=4.8$ and 6.0~K with a reduced incident neutron energy $E_\text{i}=1.3$~meV to achieve improved energy resolution. Note that the anisotropy gap decreases with increasing temperature but does not close at the commensurate-to-incommensurate phase transition ($T_1 = 4.8$~K) and remains finite within the incommensurate phase at 6.0~K.

\begin{figure}[t!]
\includegraphics[width=\columnwidth]{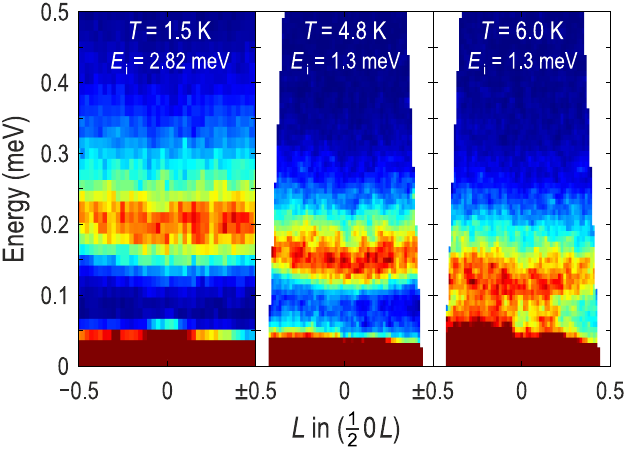}\vspace{-4pt}
\caption{Out-of-plane cuts through the TOF datasets passing through the ordering vector in the $(\frac{1}{2}0L)$ direction. The data are shown at three temperatures: 1.5, 4.8, and 6.0~K, as indicated. The data at elevated temperatures were measured with a reduced $E_\text{i}=1.3$~meV to improve the energy resolution.  Colour scale is the same as in Fig.~2 in the main text.}
\label{Fig:LET_Gap}\vspace{-4pt}
\end{figure}

\bigskip
\noindent\textbf{\it S2.~The influence of $H$-integration on the neutron data}\vspace*{5pt}

\begin{figure*}[t!]
\includegraphics[width=0.7\textwidth]{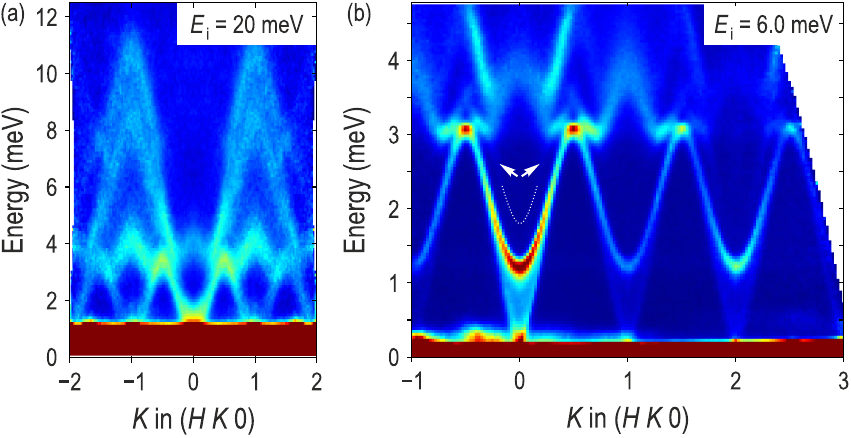}\vspace{-4pt}
\caption{Momentum-energy cuts along the $K$ direction, parallel to the chains, after a wide integration in the orthogonal $H$ direction. Panels (a) and (b) are to be compared with 
Figs.~\ref{Fig:INS}\,(c) 
and (d) in the main text, respectively. White arrows point to an additional weak magnon branch.  Colour scale is the same as in Fig.~2 in the main text.\vspace{-2pt}}
\label{Fig:LET_H-integrated}
\begin{minipage}[t]{\columnwidth}
\vspace{0pt}
\includegraphics[width=0.78\columnwidth]{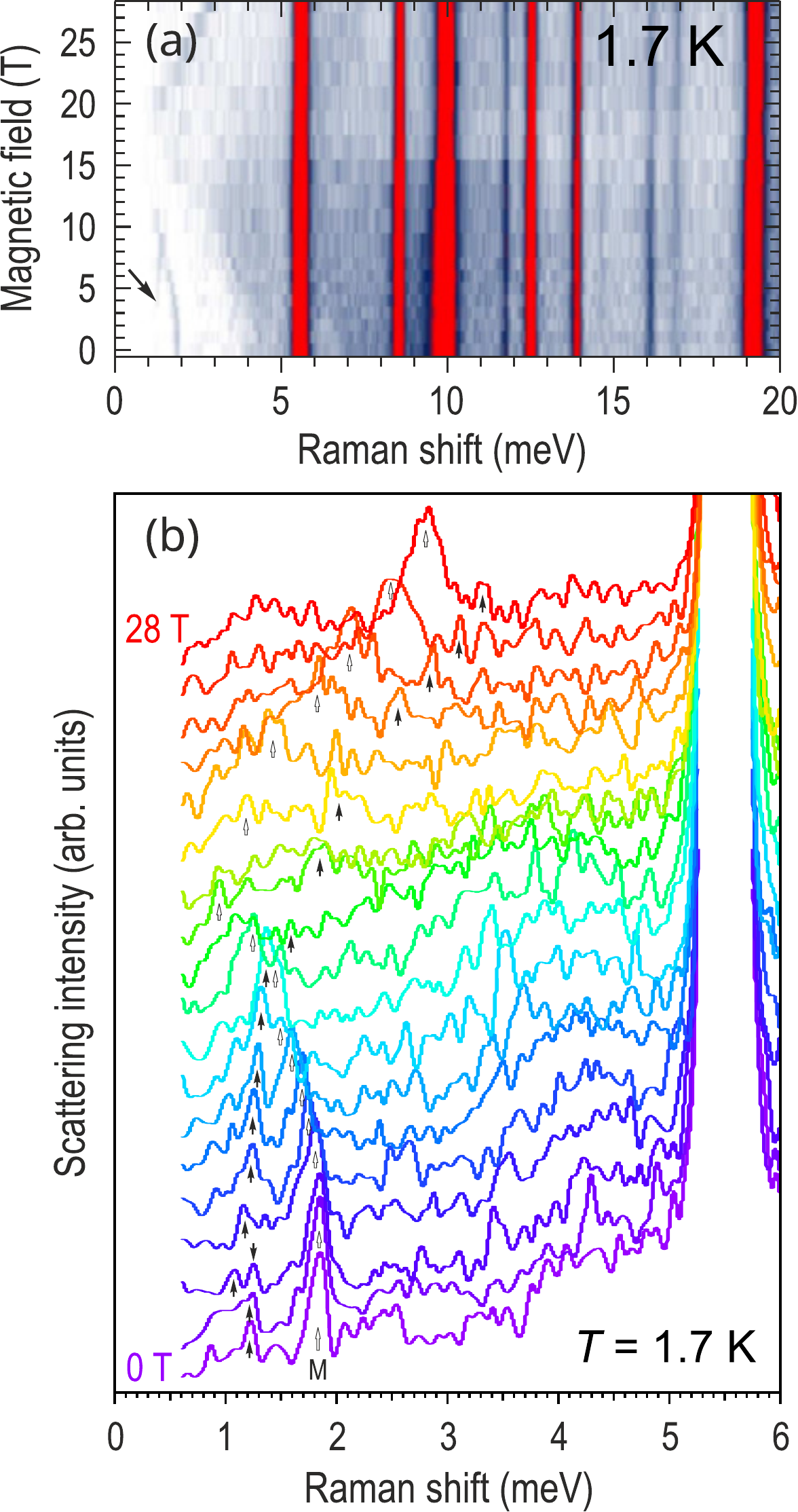}\vspace{-2pt}
\caption{Low-energy part of the 1.7-K Raman data reproduced from D.~Wulferding \textit{et al.} \cite{WulferdingIslam25}. (a)~Color map showing magnetic-field dependence of the Raman scattering intensity. The 1.2-meV magnon band is indicated by the black arrow. (b)~Low-energy part of the field dependence with the curves offset vertically for clarity (the colors are preserved from the original publication). The evolution of the two magnon peaks at 1.2 and 1.8~meV is indicated with arrows.}
\label{Fig:Raman}
\end{minipage}\hfill
\begin{minipage}[t]{\columnwidth}
\vspace{0pt}
\includegraphics[width=0.92\columnwidth]{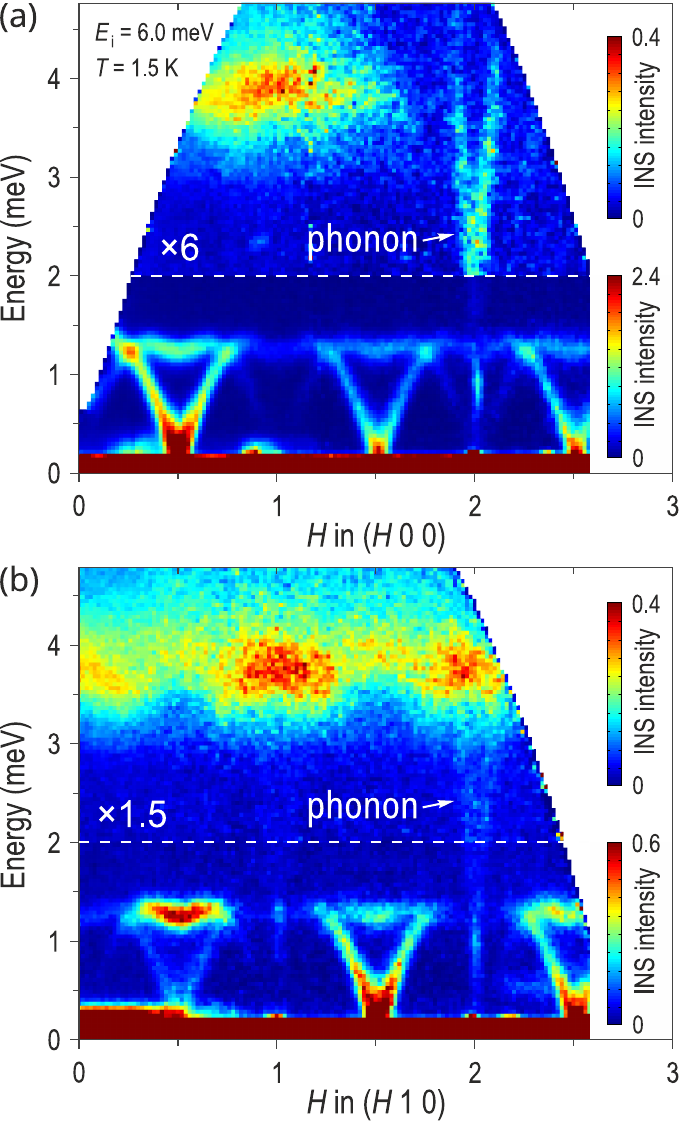}\vspace{-2pt}
\caption{Cuts through the $E_\text{i}=6.0$~meV dataset along the (a)~$(H00)$ and (b)~$(H10)$ directions that show a weak dispersion of the broad 4-meV feature in the direction orthogonal to the chains. Note that different intensity scales are used for energies above and below 2~meV (white dashed line) as indicated by the color bars. The V-shaped features emanating from $(200)$, $(110)$, and $(210)$ are acoustic phonons.}
\label{Fig:LET_Spectra_H00}
\end{minipage}
\end{figure*}

Figure~\ref{Fig:LET_H-integrated} shows cuts in the direction of the spin chains, similar to those in 
Fig.~\ref{Fig:INS}\,(c)
and (d) of the main paper, but with a wide integration in the orthogonal in-plane direction over the full accessible $H$ range. This mimics the data treatment employed by H.~Zhang \textit{et al.}~\cite{ZhangZhao20} for the analysis of the Br-botallackite data. One can see that the space between the two AFM spin-wave branches is filled with intensity that is not present in 
Fig.~\ref{Fig:INS}, 
which uses a narrow $H$-integration. This broad intensity distribution originates from the integration procedure and can be mistaken for a spinon continuum.

In Fig.~\ref{Fig:LET_H-integrated}\,(b), at higher energies between 3 and 4~meV near $K=0$, one can clearly see the difference in width between the broad 4-meV feature at the zone center and the upward-dispersing FM magnon modes to which it is connected. 

\bigskip
\noindent\textbf{\it S3.~Comparison with the Raman data of D.~Wulferding \textit{et al.}}\vspace*{5pt}

The broad $H$-integration of the $E_\text{i}=20$~meV data in Fig.~\ref{Fig:LET_H-integrated}\,(a) makes the splitting of the high-energy magnon band into two peaks at $\sim$8 and 11~meV more obvious. It also highlights weak features in the INS spectrum that are not obvious in 
Fig.~\ref{Fig:INS} 
due to limited statistics. Note, for example, a weak additional magnon band within the upper branch of the AFM magnon, indicated with a pair of white arrows. This magnon branch has a vanishing dynamical structure factor at the zone center, but by extrapolation (white dotted line) one can estimate the bottom of this band to lie at approximately 1.8~meV. This energy matches perfectly with the magnon mode (feature `M') observed in Raman spectroscopy by D.~Wulferding \textit{et al.}~\cite{WulferdingIslam25}. On the other hand, the most intense zone-center excitation that appears in our INS data at 1.2~meV has only weak intensity in Raman scattering in zero magnetic field. However, it gains intensity in an applied field, and a careful inspection of Fig.~4\,(f) in Ref.~\cite{WulferdingIslam25} indeed suggests that there is a second feature appearing in finite fields that extrapolates to 1.2~meV at $B\rightarrow0$.

In Fig.~\ref{Fig:Raman} we reproduce the low-energy portion of the Raman data from Ref.~\cite{WulferdingIslam25} that is relevant to our discussion. One can clearly see two features at approximately 1.2 and 1.8~meV that move toward each other as the magnetic field is increased and cross at approximately 8~T. The stronger, higher-energy Raman feature (termed `M' by the authors of the original paper) corresponds to the weak, barely visible INS mode [white arrows in Fig.~\ref{Fig:LET_H-integrated}\,(b)], whereas the weaker, lower-energy Raman feature that was not previously discussed corresponds to the strongest INS peak at the zone center. The bottom of the AFM magnon band at 0.2~meV lies outside of the measured energy window in the Raman spectra and therefore could not be observed. The Raman data also seem to exhibit a broad peak in the vicinity of 4~meV, which may originate from the broad 4-meV feature visible in our INS data.

This comparison emphasizes the complementarity of INS and Raman spectroscopies. Considering that the magnetic unit cell of rouaite contains 8 magnetic sites, the spin-wave spectrum may consist of up to 8 spin-wave modes, some of which may be degenerate or have vanishing dynamical structure factor in zero magnetic field.

\begin{figure}[t!]
\includegraphics[width=\columnwidth]{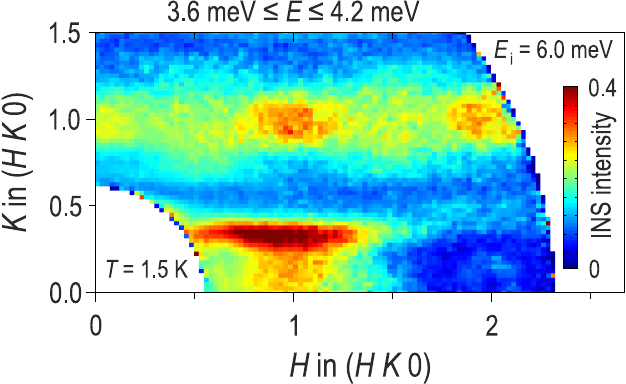}\vspace{-4pt}
\caption{Dynamical structure factor of the 4-meV feature in the $(HK0)$ plane. The INS intensity ($E_\text{i}=6.0$~meV, $T=1.5$~K) has been integrated from 3.6 to 4.2~meV in energy to cover the entire width of the peak seen in Fig.~\ref{Fig:LET_Spectra_H00}.}
\label{Fig:LET_Spectra_HK0}\vspace{-4pt}
\end{figure}

In addition to the sharp magnon modes, Raman spectra also exhibit a broad magnetic continuum with an onset at 3~meV, as reported by D.~Wulferding \textit{et al.} \cite{WulferdingIslam25}. This continuum (or at least its low-energy part) could be associated with the broad feature centered at 4~meV in our INS data. As shown in Figs.~\ref{Fig:LET_Spectra_H00} and \ref{Fig:LET_Spectra_HK0}, this feature has only a weak dispersion along the $H$ direction and has the highest intensity at the zone center, with a low-energy onset at approximately 3~meV, in agreement with the Raman measurements.  We note that similar Raman and THz measurements have very recently been reported on Cu$_2$(OH)$_3$Br, including a broad hump of intensity around 5--10\,meV\,\cite{Reinold2026A,Reinold2026B}.

\bigskip
\noindent\textbf{\it S4.~Details of DFT calculations}\vspace*{5pt}

\begin{table}[t]
\caption{\label{tab:hpos}
GGA+$U$-optimized hydrogen positions in space group $P2_1$ (No.\ 4).}
\begin{ruledtabular}
\begin{tabular}{rddd}
site & \multicolumn{1}{c}{$x/a$} & \multicolumn{1}{c}{$y/b$} & \multicolumn{1}{c}{$z/c$} \\ \hline
H(1) &  0.906445 & 0.260938 & 0.720480 \\
H(2) &  0.306186 & 0.027241 & 0.734825 \\
H(3) &  0.717943 & 0.981910 & 0.262284 \\
\end{tabular}
\end{ruledtabular}
\end{table}

\begin{figure}[t]
  \includegraphics[width=8.6cm]{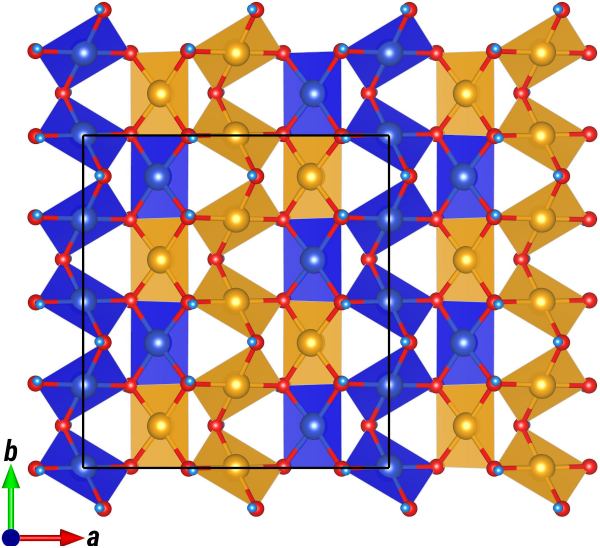}
  \caption{\label{fig:collinear_gs}The lowest-energy collinear configuration (top row in Table~\ref{tab:configs}), CuO$_4$ plaquettes with up (down) magnetic moments are shown in blue (orange). The black rectangle shows the quadrupled unit cell.}
\end{figure}

First-principles calculations were performed using the generalized gradient approximation (GGA)~\cite{PBE96} implemented in the full-potential code FPLO
version 22~\cite{FPLO}. To account for electronic interactions in the $d$ shell
of Cu, we used the GGA+$U$ functional with the Coulomb repulsion $U$ =
9.5\,eV and the Hund exchange $J_d$ = 1\,eV as interaction parameters, and
the fully localized limit as the double-counting correction.  All calculations
were performed using scalar relativistic corrections; spin-orbit coupling
has not been included.  We used the experimental crystal structure determined
by single-crystal neutron diffraction, with unit cell parameters and atomic
coordinates provided, respectively, in Tables I and IV of
Ref.~\cite{ChakkingalKulbakov24}. Since the measured samples were not fully
deuterated, sizable error bars are expected for hydrogen positions. We
optimized them with respect to the GGA+$U$ total energy (on a $k$-mesh of
$11\!\times\!10\!\times\!9$ points) in the antiferromagnetic state described by
the experimental space group $P2_1$ (No.\ 4); the refined fractional coordinates are
provided in Table~\ref{tab:hpos}.  All other structural parameters were kept
fixed. The residual forces on H positions in the resulting structure were below
0.001\,eV/\r{A}. To estimate all relevant exchange interactions, we lowered the
space group to $P1(1)$, doubled the cell along $a$ and $b$ directions, and
calculated GGA+$U$ total energies of 48 different magnetic configurations (Table~\ref{tab:configs}); in
these spin-polarized calculations, the Brillouin zone was sampled with
$2\!\times\!2\!\times\!4$ $k$-points.  Magnetic exchange interactions were
obtained by mapping the GGA+$U$ total energies onto a classical model and
performing bootstrapped least-squares fits.

\begin{table}[t]
\caption{\label{tab:cupos}Fractional coordinates of Cu atoms in the quadrupled
(doubled along $a$ and $b$) cell used for total-energy GGA+$U$ calculations.
}
\begin{ruledtabular}
\begin{tabular}{rddd}
site & \multicolumn{1}{c}{$x/a$} & \multicolumn{1}{c}{$y/b$} & \multicolumn{1}{c}{$z/c$} \\ \hline
Cu(1) & -0.0008 &  -0.0003 & -0.0071 \\
Cu(2) & -0.0008 &  0.4997 & -0.0071 \\
Cu(3) & 0.4992 &  -0.0003 & -0.0071 \\
Cu(4) & 0.4992 &  0.4997 & -0.0071 \\
Cu(5) & 0.0008 &  0.2497 & 0.0071 \\
Cu(6) & 0.0008 &  0.7497 & 0.0071 \\
Cu(7) & 0.5008 &  0.2497 & 0.0071 \\
Cu(8) & 0.5008 &  0.7497 & 0.0071 \\
Cu(9) & -0.24525 &  0.126 & -0.0013 \\
Cu(10) & -0.24525 &  0.626 & -0.0013 \\
Cu(11) & 0.25475 &  0.126 & -0.0013 \\
Cu(12) & 0.25475 &  0.626 & -0.0013 \\
Cu(13) & 0.24525 &  -0.124 & 0.0013 \\
Cu(14) & 0.24525 &  0.376 & 0.0013 \\
Cu(15) & 0.74525 &  -0.124 & 0.0013 \\
Cu(16) & 0.74525 &  0.376 & 0.0013 \\
\end{tabular}
\end{ruledtabular}
\end{table}

\begin{table*}[t]
\caption{\label{tab:configs}48 magnetic configurations and the corresponding
multipliers of exchange interactions in the classical $S=\frac12$ Heisenberg
model. For the numbering of Cu sites in the first column, see
Table~\ref{tab:cupos}.  The last three columns show the energy of the classical
$S=\frac12$ Heisenberg model computed with the solution from
Table~\ref{tab:js}, 
the GGA+$U$ total energy (with the nonmagnetic contribution
$E_0$ subtracted), and the difference $\Delta\!E$ between the two quantities.
All values are in K.
}
\begin{ruledtabular}
\begin{tabular}{rrrrrrrrrddd}
magnetic state & $J_{1a}$ & $J_{1b}$ & $J_{1d}$ & $J_{1c}$ & $J_{1f}$ & $J_{1e}$ & $J_{\text{nnn}}$ & $J_x$ & \multicolumn{1}{c}{$E$} & \multicolumn{1}{c}{$E-E_0$} & \multicolumn{1}{c}{$\Delta\!E$} \\
& & & & & & & & & \multicolumn{1}{c}{(Heisenberg)} & \multicolumn{1}{c}{(GGA+$U$)}  \\ \hline
$\downarrow\downarrow\uparrow\uparrow\downarrow\downarrow\uparrow\uparrow\downarrow\downarrow\uparrow\uparrow\downarrow\downarrow\uparrow\uparrow$ & $\phantom{+}2$ & $-2$ &$\phantom{+}2$ &$\phantom{+}2$ & $-2$ & $-2$ &$\phantom{+}2$ & $-2$ & -168.60 & -168.45 & -0.15 \\
$\downarrow\downarrow\downarrow\uparrow\downarrow\downarrow\uparrow\uparrow\uparrow\downarrow\uparrow\uparrow\downarrow\downarrow\uparrow\uparrow$ & $\phantom{+}1$ & $-1$ &$\phantom{+}1$ &$\phantom{+}1$ & $-1$ & $-2$ &$\phantom{+}1$ & $-1$ & -111.56 & -112.44 & 0.89 \\
$\downarrow\downarrow\downarrow\uparrow\downarrow\downarrow\uparrow\uparrow\downarrow\downarrow\uparrow\uparrow\uparrow\downarrow\uparrow\uparrow$ & $\phantom{+}1$ & $-1$ &$\phantom{+}1$ &$\phantom{+}1$ & $-2$ & $-1$ &$\phantom{+}1$ & $-1$ & -101.29 & -102.17 & 0.88 \\
$\downarrow\downarrow\downarrow\uparrow\downarrow\downarrow\uparrow\uparrow\downarrow\downarrow\uparrow\downarrow\downarrow\downarrow\uparrow\uparrow$ & $\phantom{+}1$ & $-1$ &$\phantom{+}1$ &$\phantom{+}1$ & $-1$ & $-1$ &$\phantom{+}1$ & $-1$ & -84.30 & -86.03 & 1.73 \\
$\downarrow\downarrow\downarrow\uparrow\downarrow\downarrow\uparrow\uparrow\downarrow\downarrow\downarrow\uparrow\downarrow\downarrow\uparrow\uparrow$ & $\phantom{+}1$ & $-1$ &$\phantom{+}1$ &$\phantom{+}2$ & $-1$ & $-1$ &$\phantom{+}1$ & $-1$ & -82.45 & -81.25 & -1.20 \\
$\downarrow\downarrow\downarrow\uparrow\downarrow\downarrow\uparrow\uparrow\downarrow\downarrow\uparrow\uparrow\downarrow\downarrow\downarrow\uparrow$ & $\phantom{+}1$ & $-1$ &$\phantom{+}2$ &$\phantom{+}1$ & $-1$ & $-1$ &$\phantom{+}1$ & $-1$ & -79.19 & -78.03 & -1.16 \\
$\downarrow\downarrow\downarrow\downarrow\downarrow\downarrow\downarrow\uparrow\uparrow\uparrow\uparrow\uparrow\downarrow\uparrow\uparrow\uparrow$ & $\phantom{+}1$ &$\phantom{+}1$ & $-1$ & $-1$ & $-1$ & $-2$ &$\phantom{+}1$ &$\phantom{+}1$ & -42.91 & -42.90 & -0.01 \\
$\downarrow\downarrow\downarrow\downarrow\downarrow\downarrow\downarrow\uparrow\uparrow\downarrow\uparrow\uparrow\uparrow\uparrow\uparrow\uparrow$ & $\phantom{+}1$ &$\phantom{+}1$ & $-1$ & $-1$ & $-2$ & $-1$ &$\phantom{+}1$ &$\phantom{+}1$ & -32.64 & -32.64 & -0.01 \\
$\downarrow\downarrow\downarrow\downarrow\downarrow\downarrow\downarrow\downarrow\uparrow\uparrow\uparrow\uparrow\uparrow\uparrow\uparrow\uparrow$ & $\phantom{+}2$ &$\phantom{+}2$ & $-2$ & $-2$ & $-2$ & $-2$ &$\phantom{+}2$ &$\phantom{+}2$ & -31.30 & -30.09 & -1.21 \\
$\downarrow\downarrow\downarrow\downarrow\downarrow\uparrow\uparrow\uparrow\uparrow\uparrow\uparrow\uparrow\downarrow\downarrow\downarrow\uparrow$ &  $-1$ & $-1$ &$\phantom{+}1$ & $-1$ &$\phantom{+}1$ & $-2$ &$\phantom{+}1$ &$\phantom{+}1$ & -24.10 & -24.53 & 0.44 \\
$\downarrow\downarrow\downarrow\downarrow\downarrow\downarrow\downarrow\uparrow\uparrow\uparrow\uparrow\downarrow\uparrow\uparrow\uparrow\uparrow$ & $\phantom{+}1$ &$\phantom{+}1$ & $-2$ & $-1$ & $-1$ & $-1$ &$\phantom{+}1$ &$\phantom{+}1$ & -20.76 & -21.30 & 0.53 \\
$\downarrow\downarrow\uparrow\uparrow\downarrow\downarrow\uparrow\uparrow\uparrow\uparrow\downarrow\downarrow\uparrow\uparrow\downarrow\downarrow$ & $\phantom{+}2$ & $-2$ & $-2$ & $-2$ &$\phantom{+}2$ &$\phantom{+}2$ &$\phantom{+}2$ & $-2$ & -19.45 & -18.59 & -0.86 \\
$\downarrow\downarrow\downarrow\downarrow\downarrow\downarrow\downarrow\uparrow\uparrow\uparrow\uparrow\uparrow\uparrow\uparrow\downarrow\uparrow$ & $\phantom{+}1$ &$\phantom{+}1$ & $-1$ & $-2$ & $-1$ & $-1$ &$\phantom{+}1$ &$\phantom{+}1$ & -17.50 & -18.03 & 0.53 \\
$\downarrow\downarrow\downarrow\downarrow\downarrow\downarrow\downarrow\uparrow\downarrow\uparrow\uparrow\uparrow\uparrow\uparrow\uparrow\uparrow$ & $\phantom{+}1$ &$\phantom{+}1$ & $-1$ & $-1$ & $-1$ & $-1$ &$\phantom{+}1$ &$\phantom{+}1$ & -15.65 & -14.05 & -1.60 \\
$\downarrow\downarrow\downarrow\uparrow\downarrow\downarrow\uparrow\uparrow\uparrow\uparrow\downarrow\downarrow\uparrow\uparrow\uparrow\downarrow$ & $\phantom{+}1$ & $-1$ & $-2$ & $-1$ &$\phantom{+}1$ &$\phantom{+}1$ &$\phantom{+}1$ & $-1$ & -14.84 & -15.90 & 1.06 \\
$\downarrow\downarrow\downarrow\uparrow\downarrow\downarrow\uparrow\uparrow\uparrow\uparrow\uparrow\downarrow\uparrow\uparrow\downarrow\downarrow$ & $\phantom{+}1$ & $-1$ & $-1$ & $-2$ &$\phantom{+}1$ &$\phantom{+}1$ &$\phantom{+}1$ & $-1$ & -11.57 & -12.62 & 1.05 \\
$\downarrow\downarrow\downarrow\uparrow\downarrow\downarrow\uparrow\uparrow\uparrow\downarrow\downarrow\downarrow\uparrow\uparrow\downarrow\downarrow$ & $\phantom{+}1$ & $-1$ & $-1$ & $-1$ &$\phantom{+}1$ &$\phantom{+}1$ &$\phantom{+}1$ & $-1$ & -9.72 & -10.58 & 0.85 \\
$\downarrow\downarrow\downarrow\downarrow\downarrow\uparrow\uparrow\uparrow\downarrow\downarrow\uparrow\downarrow\uparrow\uparrow\uparrow\uparrow$ &  $-1$ & $-1$ & $-1$ &$\phantom{+}1$ & $-2$ &$\phantom{+}1$ &$\phantom{+}1$ &$\phantom{+}1$ & 0.16 & -0.25 & 0.41 \\
$\downarrow\downarrow\downarrow\downarrow\downarrow\uparrow\uparrow\uparrow\uparrow\uparrow\uparrow\uparrow\downarrow\uparrow\downarrow\downarrow$ &  $-1$ & $-1$ &$\phantom{+}1$ & $-2$ &$\phantom{+}1$ & $-1$ &$\phantom{+}1$ &$\phantom{+}1$ & 1.31 & 0.38 & 0.93 \\
$\downarrow\downarrow\downarrow\downarrow\downarrow\uparrow\uparrow\uparrow\uparrow\downarrow\uparrow\uparrow\downarrow\downarrow\downarrow\downarrow$ &  $-1$ & $-1$ &$\phantom{+}1$ & $-1$ &$\phantom{+}1$ & $-1$ &$\phantom{+}1$ &$\phantom{+}1$ & 3.16 & 3.22 & -0.06 \\
$\downarrow\downarrow\downarrow\downarrow\uparrow\uparrow\uparrow\uparrow\uparrow\uparrow\uparrow\uparrow\downarrow\downarrow\downarrow\downarrow$ &  $-2$ & $-2$ &$\phantom{+}2$ & $-2$ &$\phantom{+}2$ & $-2$ &$\phantom{+}2$ &$\phantom{+}2$ & 6.32 & 5.82 & 0.50 \\
$\downarrow\downarrow\downarrow\uparrow\downarrow\downarrow\uparrow\uparrow\uparrow\uparrow\downarrow\downarrow\downarrow\uparrow\downarrow\downarrow$ & $\phantom{+}1$ & $-1$ & $-1$ & $-1$ &$\phantom{+}2$ &$\phantom{+}1$ &$\phantom{+}1$ & $-1$ & 7.27 & 8.33 & -1.06 \\
$\downarrow\downarrow\downarrow\downarrow\downarrow\uparrow\uparrow\uparrow\downarrow\uparrow\uparrow\uparrow\downarrow\downarrow\downarrow\downarrow$ &  $-1$ & $-1$ &$\phantom{+}2$ & $-1$ &$\phantom{+}1$ & $-1$ &$\phantom{+}1$ &$\phantom{+}1$ & 8.27 & 9.64 & -1.37 \\
$\downarrow\downarrow\downarrow\downarrow\downarrow\uparrow\uparrow\uparrow\uparrow\downarrow\downarrow\downarrow\uparrow\uparrow\uparrow\uparrow$ &  $-1$ & $-1$ & $-2$ &$\phantom{+}1$ & $-1$ &$\phantom{+}1$ &$\phantom{+}1$ &$\phantom{+}1$ & 12.04 & 11.13 & 0.92 \\
$\downarrow\downarrow\downarrow\downarrow\downarrow\uparrow\uparrow\uparrow\downarrow\downarrow\downarrow\downarrow\downarrow\uparrow\uparrow\uparrow$ &  $-1$ & $-1$ & $-1$ &$\phantom{+}1$ & $-1$ &$\phantom{+}1$ &$\phantom{+}1$ &$\phantom{+}1$ & 17.15 & 17.25 & -0.09 \\
$\downarrow\downarrow\downarrow\uparrow\downarrow\downarrow\uparrow\uparrow\downarrow\uparrow\downarrow\downarrow\uparrow\uparrow\downarrow\downarrow$ & $\phantom{+}1$ & $-1$ & $-1$ & $-1$ &$\phantom{+}1$ &$\phantom{+}2$ &$\phantom{+}1$ & $-1$ & 17.53 & 18.54 & -1.01 \\
$\downarrow\downarrow\downarrow\downarrow\downarrow\uparrow\uparrow\uparrow\downarrow\downarrow\downarrow\downarrow\uparrow\downarrow\uparrow\uparrow$ &  $-1$ & $-1$ & $-1$ &$\phantom{+}2$ & $-1$ &$\phantom{+}1$ &$\phantom{+}1$ &$\phantom{+}1$ & 19.00 & 20.39 & -1.39 \\
$\downarrow\downarrow\downarrow\downarrow\downarrow\uparrow\uparrow\uparrow\uparrow\uparrow\downarrow\uparrow\downarrow\downarrow\downarrow\downarrow$ &  $-1$ & $-1$ &$\phantom{+}1$ & $-1$ &$\phantom{+}2$ & $-1$ &$\phantom{+}1$ &$\phantom{+}1$ & 20.15 & 21.09 & -0.93 \\
$\downarrow\downarrow\downarrow\downarrow\uparrow\uparrow\uparrow\uparrow\downarrow\downarrow\downarrow\downarrow\uparrow\uparrow\uparrow\uparrow$ &  $-2$ & $-2$ & $-2$ &$\phantom{+}2$ & $-2$ &$\phantom{+}2$ &$\phantom{+}2$ &$\phantom{+}2$ & 34.31 & 33.80 & 0.51 \\
$\downarrow\downarrow\downarrow\downarrow\downarrow\uparrow\uparrow\uparrow\downarrow\downarrow\downarrow\downarrow\uparrow\uparrow\uparrow\downarrow$ &  $-1$ & $-1$ & $-1$ &$\phantom{+}1$ & $-1$ &$\phantom{+}2$ &$\phantom{+}1$ &$\phantom{+}1$ & 44.41 & 45.28 & -0.87 \\
$\downarrow\downarrow\downarrow\uparrow\uparrow\uparrow\downarrow\downarrow\uparrow\downarrow\uparrow\uparrow\uparrow\uparrow\downarrow\downarrow$ &  $-1$ &$\phantom{+}1$ & $-1$ &$\phantom{+}1$ &$\phantom{+}1$ & $-2$ &$\phantom{+}1$ & $-1$ & 55.73 & 55.58 & 0.16 \\
$\downarrow\downarrow\downarrow\uparrow\uparrow\uparrow\downarrow\downarrow\downarrow\downarrow\uparrow\uparrow\uparrow\uparrow\uparrow\downarrow$ &  $-1$ &$\phantom{+}1$ & $-2$ &$\phantom{+}1$ &$\phantom{+}1$ & $-1$ &$\phantom{+}1$ & $-1$ & 77.88 & 77.20 & 0.67 \\
$\downarrow\downarrow\downarrow\uparrow\uparrow\uparrow\downarrow\downarrow\downarrow\downarrow\uparrow\downarrow\uparrow\uparrow\downarrow\downarrow$ &  $-1$ &$\phantom{+}1$ & $-1$ &$\phantom{+}1$ &$\phantom{+}1$ & $-1$ &$\phantom{+}1$ & $-1$ & 82.99 & 81.33 & 1.66 \\
$\downarrow\downarrow\downarrow\uparrow\uparrow\uparrow\downarrow\downarrow\downarrow\downarrow\downarrow\uparrow\uparrow\uparrow\downarrow\downarrow$ &  $-1$ &$\phantom{+}1$ & $-1$ &$\phantom{+}2$ &$\phantom{+}1$ & $-1$ &$\phantom{+}1$ & $-1$ & 84.84 & 85.96 & -1.12 \\
$\downarrow\downarrow\downarrow\downarrow\downarrow\downarrow\downarrow\uparrow\downarrow\downarrow\downarrow\downarrow\downarrow\downarrow\downarrow\uparrow$ & $\phantom{+}1$ &$\phantom{+}1$ &$\phantom{+}1$ &$\phantom{+}1$ &$\phantom{+}1$ &$\phantom{+}1$ &$\phantom{+}1$ &$\phantom{+}1$ & 86.77 & 85.21 & 1.56 \\
$\downarrow\downarrow\downarrow\downarrow\downarrow\downarrow\downarrow\uparrow\downarrow\downarrow\downarrow\downarrow\downarrow\downarrow\uparrow\downarrow$ & $\phantom{+}1$ &$\phantom{+}1$ &$\phantom{+}1$ &$\phantom{+}2$ &$\phantom{+}1$ &$\phantom{+}1$ &$\phantom{+}1$ &$\phantom{+}1$ & 88.62 & 89.41 & -0.79 \\
$\downarrow\downarrow\downarrow\downarrow\downarrow\downarrow\downarrow\uparrow\downarrow\downarrow\downarrow\uparrow\downarrow\downarrow\downarrow\downarrow$ & $\phantom{+}1$ &$\phantom{+}1$ &$\phantom{+}2$ &$\phantom{+}1$ &$\phantom{+}1$ &$\phantom{+}1$ &$\phantom{+}1$ &$\phantom{+}1$ & 91.88 & 92.65 & -0.77 \\
$\downarrow\downarrow\downarrow\uparrow\uparrow\uparrow\downarrow\downarrow\uparrow\uparrow\downarrow\downarrow\uparrow\downarrow\uparrow\uparrow$ &  $-1$ &$\phantom{+}1$ &$\phantom{+}1$ & $-1$ & $-2$ &$\phantom{+}1$ &$\phantom{+}1$ & $-1$ & 93.05 & 92.83 & 0.23 \\
$\downarrow\downarrow\downarrow\uparrow\uparrow\uparrow\downarrow\downarrow\downarrow\downarrow\uparrow\uparrow\downarrow\uparrow\downarrow\downarrow$ &  $-1$ &$\phantom{+}1$ & $-1$ &$\phantom{+}1$ &$\phantom{+}2$ & $-1$ &$\phantom{+}1$ & $-1$ & 99.98 & 100.57 & -0.59 \\
$\downarrow\downarrow\downarrow\downarrow\downarrow\downarrow\downarrow\uparrow\downarrow\uparrow\downarrow\downarrow\downarrow\downarrow\downarrow\downarrow$ & $\phantom{+}1$ &$\phantom{+}1$ &$\phantom{+}1$ &$\phantom{+}1$ &$\phantom{+}2$ &$\phantom{+}1$ &$\phantom{+}1$ &$\phantom{+}1$ & 103.76 & 104.15 & -0.39 \\
$\downarrow\downarrow\downarrow\uparrow\uparrow\uparrow\downarrow\downarrow\uparrow\uparrow\uparrow\downarrow\downarrow\downarrow\uparrow\uparrow$ &  $-1$ &$\phantom{+}1$ &$\phantom{+}1$ & $-2$ & $-1$ &$\phantom{+}1$ &$\phantom{+}1$ & $-1$ & 108.20 & 107.50 & 0.70 \\
$\downarrow\downarrow\downarrow\uparrow\uparrow\uparrow\downarrow\downarrow\uparrow\downarrow\downarrow\downarrow\downarrow\downarrow\uparrow\uparrow$ &  $-1$ &$\phantom{+}1$ &$\phantom{+}1$ & $-1$ & $-1$ &$\phantom{+}1$ &$\phantom{+}1$ & $-1$ & 110.05 & 108.97 & 1.08 \\
$\downarrow\downarrow\downarrow\downarrow\downarrow\downarrow\downarrow\uparrow\downarrow\downarrow\downarrow\downarrow\uparrow\downarrow\downarrow\downarrow$ & $\phantom{+}1$ &$\phantom{+}1$ &$\phantom{+}1$ &$\phantom{+}1$ &$\phantom{+}1$ &$\phantom{+}2$ &$\phantom{+}1$ &$\phantom{+}1$ & 114.02 & 114.34 & -0.31 \\
$\downarrow\downarrow\downarrow\uparrow\uparrow\uparrow\downarrow\downarrow\uparrow\uparrow\downarrow\downarrow\downarrow\downarrow\downarrow\uparrow$ &  $-1$ &$\phantom{+}1$ &$\phantom{+}2$ & $-1$ & $-1$ &$\phantom{+}1$ &$\phantom{+}1$ & $-1$ & 115.16 & 116.21 & -1.05 \\
$\downarrow\downarrow\downarrow\uparrow\uparrow\uparrow\downarrow\downarrow\downarrow\uparrow\downarrow\downarrow\downarrow\downarrow\uparrow\uparrow$ &  $-1$ &$\phantom{+}1$ &$\phantom{+}1$ & $-1$ & $-1$ &$\phantom{+}2$ &$\phantom{+}1$ & $-1$ & 137.30 & 137.80 & -0.50 \\
$\downarrow\downarrow\uparrow\uparrow\uparrow\uparrow\downarrow\downarrow\downarrow\downarrow\uparrow\uparrow\uparrow\uparrow\downarrow\downarrow$ &  $-2$ &$\phantom{+}2$ & $-2$ &$\phantom{+}2$ &$\phantom{+}2$ & $-2$ &$\phantom{+}2$ & $-2$ & 165.98 & 166.34 & -0.36 \\
$\downarrow\downarrow\downarrow\downarrow\downarrow\downarrow\downarrow\downarrow\downarrow\downarrow\downarrow\downarrow\downarrow\downarrow\downarrow\downarrow$ & $\phantom{+}2$ &$\phantom{+}2$ &$\phantom{+}2$ &$\phantom{+}2$ &$\phantom{+}2$ &$\phantom{+}2$ &$\phantom{+}2$ &$\phantom{+}2$ & 173.54 & 171.70 & 1.84 \\
$\downarrow\downarrow\uparrow\uparrow\uparrow\uparrow\downarrow\downarrow\uparrow\uparrow\downarrow\downarrow\downarrow\downarrow\uparrow\uparrow$ &  $-2$ &$\phantom{+}2$ &$\phantom{+}2$ & $-2$ & $-2$ &$\phantom{+}2$ &$\phantom{+}2$ & $-2$ & 220.09 & 220.35 & -0.26 \\
\end{tabular}
\end{ruledtabular}
\end{table*}

\begin{figure*}
    \includegraphics[width=\textwidth]{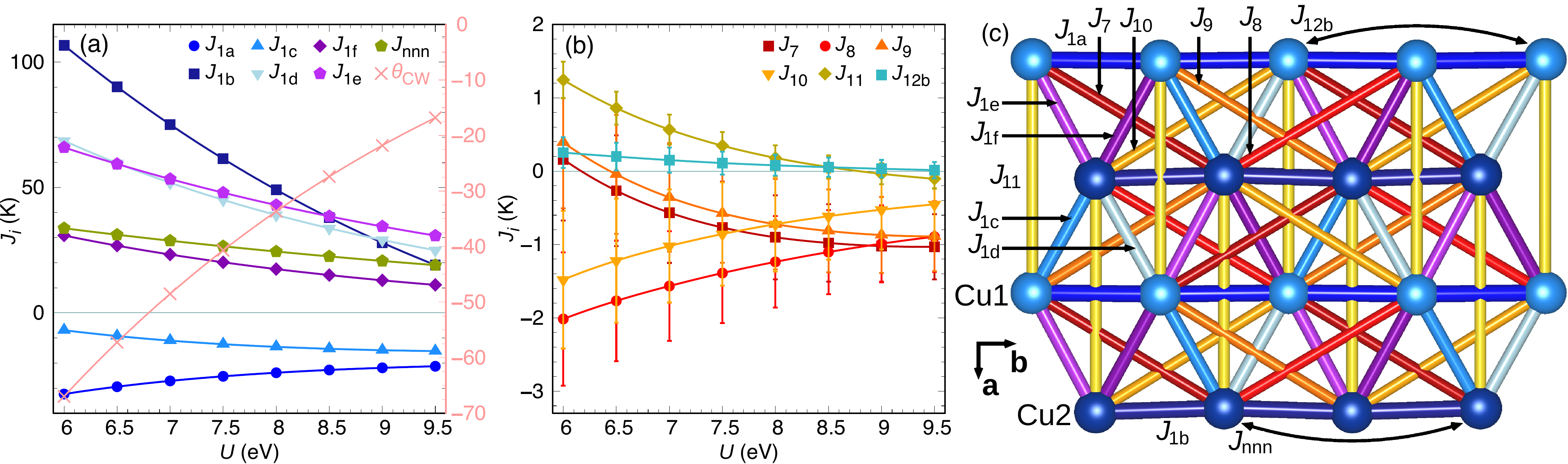}
    \caption{\label{Harald}$J$ parameters based on powder-diffraction data, as a function of the on-site repulsion $U$, with the exchanges $J_i$ identified in (c).  The resulting Curie-Weiss temperature is shown on the right axis in (a).  Our experimentally determined Curie-Weiss temperatures are slightly positive while those from Ref.~\cite{Yuan2022} were slightly negative.  This calculation indicates that small Curie-Weiss temperatures require a very high $U$.}
\end{figure*}

\bigskip
\noindent\textbf{\it S5.~DFT calculations for the NPD structure}\vspace*{5pt}

DFT+$U$ calculations were also performed based on our refined crystal structure at 10\,K from neutron powder diffraction on the HRPT diffractometer\,\cite{ChakkingalKulbakov24}.  The exchanges are shown as a function of $U$ in Fig.~\ref{Harald}, and the resulting Curie-Weiss temperature is plotted on the right vertical axis.  Our experimentally determined Curie-Weiss temperatures are smaller than those in Fig.~\ref{Harald} and positive, while those from Ref.~\cite{Yuan2022} were smaller and negative.  This suggests that a relatively large $U$ is required to model the exchanges, as was also found for the single-crystal-diffraction structure.  Longer-range exchanges (5.3--6.2\,\AA) are quite small.

We note that several exchanges (most notably $J_\text{1a}$, $J_\text{1c}$ and $J_\text{1d}$) differ very substantially in this calculation from the values in Table~\ref{tab:js}.  Both sets of DFT calculations were performed using the same code starting from high-quality refinements of neutron data taken at the same temperature on our samples, and in both cases the hydrogen positions were relaxed.  The calculations we report in the main text were based on single-crystal data, which generally provides higher accuracy for atomic positions.  Evidently, very subtle changes in the atomic positions are able to produce large, qualitative changes in the exchange parameters. This suggests that the magnetic structure is likely highly sensitive to small perturbations, and we would expect significant changes at relatively low pressure or strain.

\bigskip
\noindent\textbf{\it S6.~Dynamical density-matrix renormalization group}\vspace*{5pt}

We employ the dynamical density-matrix renormalization group
(DDMRG) method~\cite{Jeckelmann02} to calculate the zero-temperature
dynamical spin structure factor of the spin Hamiltonian introduced
in the main text. For a spin component $\alpha$, we define
\begin{equation}
	S^{\alpha\alpha}(\mathbf Q,\omega)
	=-\frac{1}{\pi}\operatorname{Im}
	\left\langle 0\left|
	S^\alpha_{-\mathbf Q}
	\frac{1}{\omega+E_0-\hat H+\mathrm{i}\eta}
	S^\alpha_{\mathbf Q}
	\right|0\right\rangle,
	\label{eq:ddmrg_sqw}
\end{equation}
where $|0\rangle$ and $E_0$ are the ground state and its energy,
respectively, and
\begin{equation}
	S^\alpha_{\mathbf Q}
	=\frac{1}{\sqrt{N}}\sum_j
	e^{\mathrm{i}\mathbf Q\cdot\mathbf r_j}S_j^\alpha.
	\label{eq:ddmrg_fourier}
\end{equation}
Here, $N$ is the number of spins and $\mathbf r_j$ is the position
of site $j$. In the following, we evaluate the $\alpha=z$ component
and denote $S^{zz}(\mathbf Q,\omega)$ by $S(\mathbf Q,\omega)$.
The positive parameter $\eta$ is the half width at half maximum
of the Lorentzian broadening. Equivalently,
\begin{equation}
	S^{\alpha\alpha}(\mathbf Q,\omega)
	=\sum_n
	\left|\langle n|S^\alpha_{\mathbf Q}|0\rangle\right|^2
	\frac{\eta/\pi}{[\omega-(E_n-E_0)]^2+\eta^2},
	\label{eq:ddmrg_lehmann}
\end{equation}
where $|n\rangle$ denotes an eigenstate of $\hat H$ with energy
$E_n$. This response includes both sharp modes and spectral
weight distributed over multiple excited states.

To evaluate the Fourier transform using real-valued source
vectors, we introduce the cosine and sine operators
\begin{align}
	C^\alpha_{\mathbf Q}
	&=\frac{1}{\sqrt{N}}\sum_j
	\cos(\mathbf Q\cdot\mathbf r_j)S_j^\alpha,\\
	D^\alpha_{\mathbf Q}
	&=\frac{1}{\sqrt{N}}\sum_j
	\sin(\mathbf Q\cdot\mathbf r_j)S_j^\alpha.
\end{align}
When the Hamiltonian and these operators have real matrix
representations and the ground state is chosen real, the
cosine--sine interference terms cancel in the spectral function.
The structure factor can then be obtained by adding the two
responses calculated separately:
\begin{align}
	\nonumber
	S^{\alpha\alpha}(\mathbf Q,\omega)
	=\sum_n&\left[
	|\langle n|C^\alpha_{\mathbf Q}|0\rangle|^2
	+|\langle n|D^\alpha_{\mathbf Q}|0\rangle|^2
	\right]\\
	&\times \frac{\eta/\pi}{[\omega-(E_n-E_0)]^2+\eta^2}.
	\label{eq:ddmrg_cos_sin}
\end{align}
Along the open chain direction, this Fourier transform can be
evaluated at arbitrary wave vectors, although the intrinsic
momentum resolution remains limited by the finite cluster length.

\begin{figure*}
	\includegraphics[width=\textwidth]{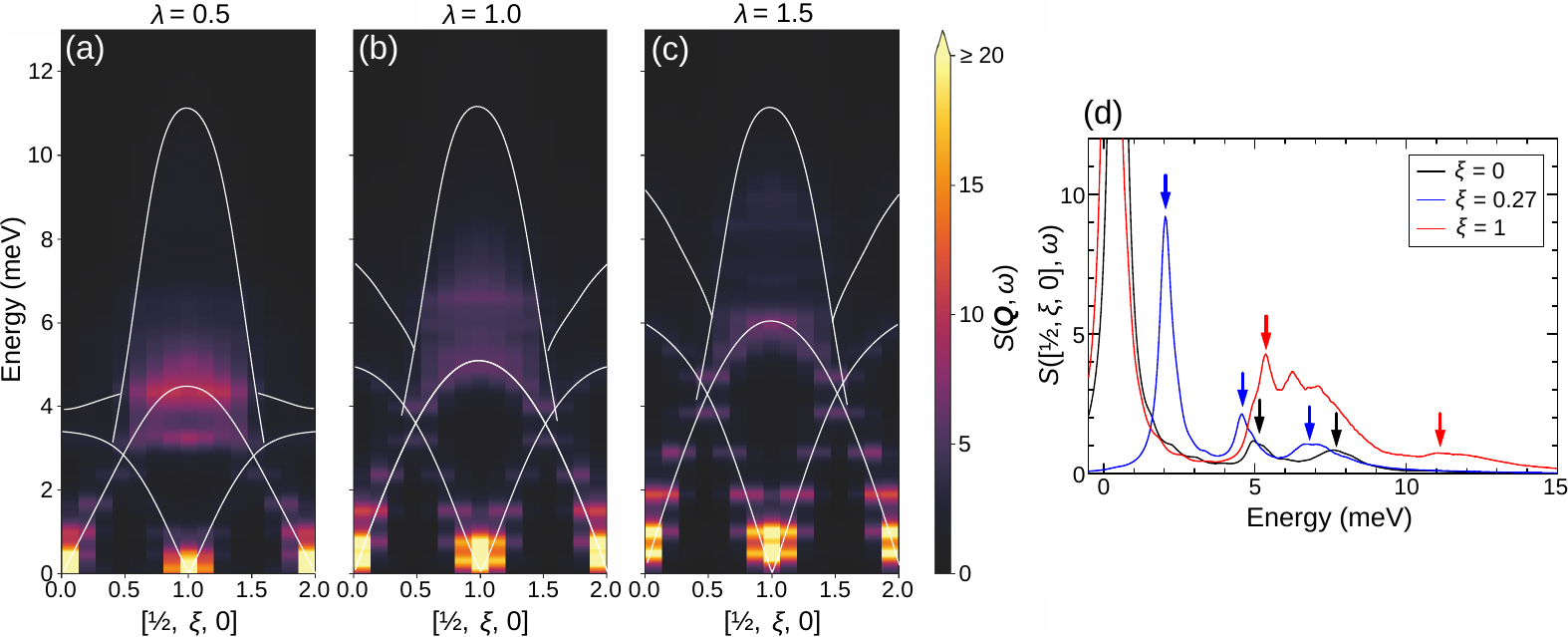}\vspace{-4pt}
\caption{DDMRG results for the dynamical spin structure factor.
	(a)--(c) Momentum--energy maps calculated on a $9\times4$
	cluster along $\mathbf Q=(\frac12,\xi,0)$ over
	$0\leq\xi\leq2$ for $\lambda=0.5$, $1.0$, and $1.5$,
	respectively.
	The four ratios $J_{1\mu}/J_{1a}$, with $\mu=c,d,e,f$,
	are uniformly scaled by $\lambda$ relative to the DFT+$U$
	estimates at $U=9.5$~eV. For each value of $\lambda$, the overall
	energy scale is adjusted to reproduce the experimental bandwidth
	of approximately 11~meV.
	White lines trace the excitation branches and estimated continuum
	boundaries identified from the energy-resolved spectra.
	A common color scale is used in (a)--(c), with saturation to
	emphasize weak features.
	(d) Representative energy-resolved spectra calculated on a
	$17\times4$ cluster for $\lambda=1.0$ at
	$\xi=0$, $0.27$, and $1$.
	Arrows indicate selected peak positions and estimated continuum
	boundaries, illustrating the spectral features traced by the
	white lines in (a)--(c).
	The white crosses in Figs.~4(c) and 4(d) of the main text
	identify the corresponding types of spectral features.}
	\label{Fig:icdep}\vspace{-4pt}
\end{figure*}

We use cylindrical clusters with open boundary conditions along
the chain direction ($x$) and periodic boundary conditions in
the transverse direction ($y$).
The spectra along $\mathbf Q=(\frac12,\xi,0)$ in the main text
are calculated on a $17\times4$ cluster, whereas the
interchain-coupling comparison in Fig.~\ref{Fig:icdep} uses
$9\times4$ clusters for $\lambda=0.5$, $1.0$, and $1.5$.

The ground state is first optimized by DMRG, retaining up to
$\chi_{\rm GS}=2000$ states, and subsequently used as the
starting state for the variational DDMRG calculation at each
frequency, retaining up to $\chi_{\rm dyn}=600$ states.
We use a Lorentzian broadening with a half width at half maximum
of $\eta=0.2$~meV. The maximum discarded weight is of the order
of $10^{-4}$.

\bigskip
\noindent\textbf{\it S7.~Dependence of the excitation spectrum on interchain couplings}\vspace*{5pt}

As noted in the main text, the calculated broad continuum-like feature
appears at a higher energy ($\sim6$~meV) than observed experimentally
($\sim4$~meV). We examine the
dependence of the spectrum on the interchain coupling strength relative to the dominant
intrachain exchange $J_{1a}$ to explore whether this discrepancy can be reduced by
modifying the interchain exchanges and to see how the feature associated with vestiges of spinons evolves as the dimensionality is tuned.

To limit the number of adjustable parameters, we scale the four
interchain coupling ratios by a common factor $\lambda$:
\[
\frac{J_{1\mu}}{J_{1a}}
=
\lambda
\left(\frac{J_{1\mu}}{J_{1a}}\right)_{\mathrm{DFT}+U},
\qquad \mu=c,d,e,f,
\]
where the reference ratios are obtained from DFT+$U$ at $U=9.5$~eV,
while all other couplings are kept at their DFT+$U$ ratios to $J_{1a}$.
This preserves the relative strengths and signs of the four interchain
couplings.

Figures~\ref{Fig:icdep}(a)--(c) show the momentum-resolved
dynamical spin structure factor along $(\frac12\,\xi\,0)$ for $\lambda=0.5$, $1.0$, and $1.5$, calculated on a $9\times4$ cluster with the same broadening as in Fig.~4(c) of the main text. For each value of $\lambda$, the overall
energy scale is adjusted to reproduce the experimental bandwidth of
approximately 11~meV. Compared with the reference result at $\lambda=1$,
the broad feature near $\xi=1$ shifts to lower energies for
$\lambda=0.5$, placing its main spectral weight near 4~meV and improving
agreement with experiment.

Figure~\ref{Fig:icdep}(d) shows representative energy-resolved
spectra calculated on the $17\times4$ cluster for $\lambda=1.0$.
The arrows mark selected peak positions and estimated continuum
boundaries at $\xi=0$, $0.27$, and $1$, illustrating the procedure
used to identify the white crosses in Figs.~4(c) and 4(d) of the
main text. The same procedure is applied to the $9\times4$
results for each value of $\lambda$ to obtain the white lines
in Figs.~\ref{Fig:icdep}(a)--(c).

Although this comparison does not constitute a precise fit to the full
spectrum, it shows that the energy of the broad feature is sensitive to
the interchain coupling ratios even after matching the overall
bandwidth. Within this restricted parameter variation,
$\lambda\simeq0.5$ approximately reproduces the experimental energy of
the broad feature, suggesting that the couplings between neighboring
chains may be weaker relative to $J_{1a}$ than estimated by DFT+$U$.

Increasing the interchain coupling strength not only raises the energy of the broad feature, it also significantly reduces its intensity.  This suggests that its spectral weight is progressively transferred to magnons as the material is tuned from more one-dimensional to more two-dimensional.

\end{document}